\def\year{2019}\relax
\documentclass[letterpaper]{article} 
\usepackage{aaai19}  
\usepackage{times}  
\usepackage{helvet}  
\usepackage{courier}  
\usepackage{url}  
\usepackage{graphicx}  
\begin{document}
%
\title{Reagent: Converting Ordinary Webpages into Interactive Software Agents}
\author{Matthew Peveler\\
Rensselaer Polytechnic Institute\\
Troy, NY 12180, USA\\
pevelm@rpi.edu
\And
Jeffrey O. Kephart \and Hui Su\\
IBM Thomas J Watson Research Center\\
Yorktown Heights, NY 10598, USA\\
\{kephart,huisuibmres\}@us.ibm.com
}
\maketitle
\begin{abstract}
We introduce \textit{Reagent}, a technology that readily converts ordinary webpages containing structured data into software agents with which one can interact naturally, via a combination of speech and pointing. Previous efforts to make webpage content manipulable by third-party software components in browsers or desktop applications have generally relied upon specialized instrumentation included in the webpages -- a practice that neither scales well nor applies to pre-existing webpages. In contrast, \textit{Reagent} automatically captures semantic details and  semantically-meaningful mouse events from arbitrary webpages that contain no pre-existing special instrumentation. \textit{Reagent} combines these events with text transcriptions of user speech to derive and execute parameterized commands representing human intent. Thus, users may request various visualization or analytic operations to be performed on data displayed on a page by speaking to it and/or pointing to elements within it. When unable to infer translations between event labels and human terminology, \textit{Reagent} proactively asks users for definitions and adds them to its dictionary. We demonstrate \textit{Reagent} in the context of a collection of pre-existing webpages that contain football team and player statistics.
\end{abstract}

\section{Introduction}\label{Introduction}

With the growing ubiquity of voice-activated assistants like Siri and Alexa, society is 
quickly acclimating to interacting with AI as we do with our fellow humans. Now under development is
a new generation of more sophisticated assistants designed to help scientists, business users, and students with
cognitive tasks such as data exploration and analysis~\cite{kephart2018cognitive}, decision making~\cite{farrell2016symbiotic}, and learning language.

Practically all of these cognitive applications entail some sort of interaction with data that
is based on multi-modal inputs that include speech, pointing and gesture. To date, the assistants have supported such interactions by reading data from a database and displaying it in the form of tables or plots through web pages that are specially instrumented to capture pointing events. A stream of utterances is converted into a text stream by a speech-to-text engine and combined with the stream of pointing events to derive user intent, i.e. a parameterized command that is then executed by orchestrating one or more services and rendering the output on a display and optionally as synthesized speech played through a speaker. A major bottleneck in the creation of such cognitive assistants is that creating specially-instrumented web pages is labor-intensive. Unless this bottleneck can be removed, it seems unlikely that multi-modal cognitive assistants will become anywhere near as pervasive as the present generation of less-sophisticated bots.

In this paper, we describe \textit{Reagent}, a technology that reduces this bottleneck by readily converting ordinary non-instrumented webpages that contain structured data into software agents with which one can interact naturally, via a combination of speech and pointing. \textit{Reagent} combines streams of semantically-meaningful mouse events with speech transcriptions to derive and execute parameterized commands that represent user 
requests to visualize, extract, query, sort, filter, analyze, or otherwise manipulate data displayed on webpages. Command execution entails displaying the requested information in the original webpage or in a dynamically-constructed one, as well as playing synthesized speech. \textit{Reagent} automatically infers mappings from event labels to human-friendly terminology, or when necessary learns them actively from the user. Now we shall proceed to describe briefly how \textit{Reagent} works and illustrate its use in the context of a collection of pre-existing webpages that contain football team and player statistics.

\section{Technical Details}

\begin{figure}
\centering
\includegraphics[width=0.45\textwidth]{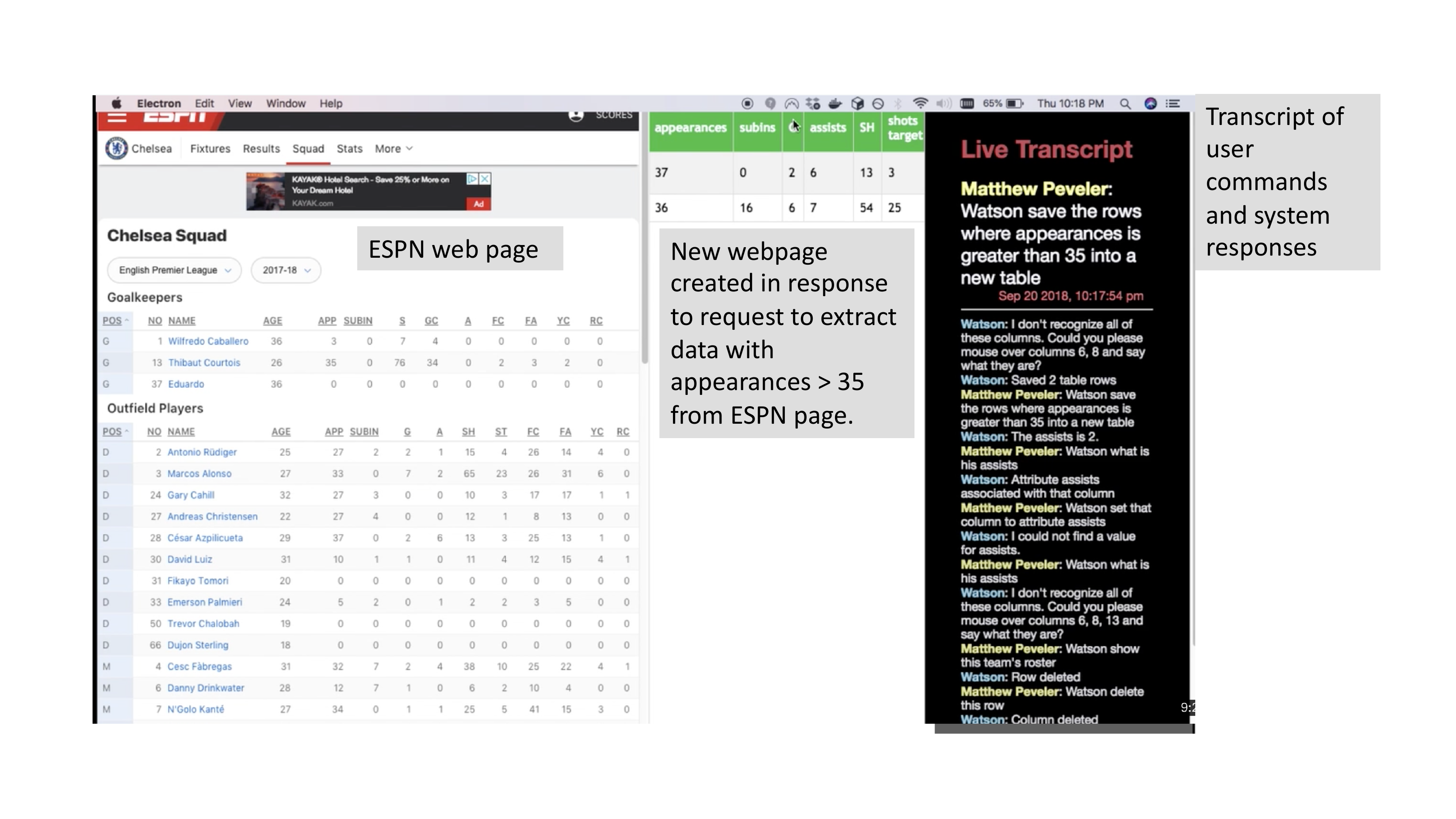}
\caption{Electron environment containing ESPN page, transcript window, and auto-generated webpage.}
\label{fig:screenshot}
\end{figure}

\textit{Reagent} is built upon Electron\footnote{https://electronjs.org/}, a node.js framework for developing desktop applications from web technologies. As illustrated in Fig.~\ref{fig:screenshot}, content is displayed on a computer screen as a collection of webpages (termed webviews in Electron).

When a user opens a webpage, \textit{Reagent} inserts a transparent layer on top of that page's webview, semantically parses the page to identify key HTML structures with which a user might interact (e.g. a table or a plot), and uses this information to bind appropriate event handlers to HTML elements, thereby preparing them to listen for user interaction events. More specifically, 
\begin{enumerate}
    \item Electron pre-loads the \textit{Reagent} system on top of the webview, and injects a websocket connection allowing it to communicate with the open webview and receive JSON-formatted events into an event buffer.
    \item Electron emits a ``DOM Ready" event to signal that the page has finished loading, whereupon \textit{Reagent} scans the page to see if it can detect any major HTML elements it should parse and bind to (e.g. a table).
    \item \textit{Reagent} injects additional code specific to the detected HTML elements that binds event listeners to these elements as well as assigning each a UUID --- an approach that makes \textit{Reagent} readily extensible to new types of elements. (We have implemented domain-independent layers designed for general tables and for plot.ly plots.)
    \item Next, \textit{Reagent} creates a MutationObserver\footnote{https://mzl.la/1exU78d} to watch for any changes to the page. If changes are detected, it performs any necessary re-injections and re-bindings to ensure that all relevant elements remain instrumented.
\end{enumerate}

Users interact with a web page by pointing to its elements (via mouse or other pointing device) and either typing into a chat window or speaking (in which case we use Watson Speech-to-Text to transcribe this to text). \textit{Reagent} calls the IBM Watson Assistant service to obtain an ``intent'' classification plus a set of entities. A parser then processes the text, the intent class, and the entities in an effort to extract a set of parameters associated with the intent. If any required parameters remain unspecified, the parser calls an API on the \textit{Reagent} event buffer to retrieve the most recent events that could possibly be mapped to the missing parameter(s), and from this generates a fully specified JSON command. The command is then executed, the result of which is displayed on the Electron display canvas, possibly accompanied by synthesized speech~\cite{divekar2018}.

Fig.~\ref{fig:screenshot} illustrates the state of the display after the user has opened an ESPN football team roster webpage and asked the system ``Show in a new table rows where \textit{appearances} is greater than 35.'' Equivalently, the user might have asked ``Show in a new table rows with this column greater than this.'' while pointing to a cell under the column `APP' with value 35. For a fuller set of capabilities that includes querying, sorting, and simple analytics like averaging, see the demo video at \url{https://bit.ly/2OGKvez}.

Frequently, webpage developers use abbreviations or synonyms that don't correspond directly to the terms that users would naturally use to interact with those pages. For example, in Fig.~\ref{fig:screenshot}, the column for ``appearances'' was labelled ``APP''. In the course of processing HTML elements, \textit{Reagent} identifies tags that may indicate more human-friendly terms, such as tooltips that reveal explanatory text when a user hovers over an element, and uses approximate text matching to infer likely associations. \textit{Reagent} can also exploit a large body of accessibility work including the W3C Standard Web Content Accessibility Guidelines\footnote{https://www.w3.org/WAI/standards-guidelines/wcag/} 
or a Voluntary Product Accessibility Template\footnote{https://www.section508.gov/sell/vpat} to automatically derive meaningful semantic information in ``hidden'' attributes. In cases where the system is unable to disambiguate the semantic information, it solicits a definition from the user. For example, with the ESPN table, the system would ask the user (via synthesized voice) to mouse over ``the column labeled \textit{A}'' and state what is the attribute. Upon hearing ``Watson assign attribute \textit{assists} to this column", \textit{Reagent} stores the mapping in a dictionary, whereupon it becomes available to any subsequently accessed bound elements on the same webpage or host site.

\section{Conclusions}

Taking advantage of commonly-occurring structural motifs and human-friendly tagging such as tooltips, \textit{Reagent} makes it easy for developers to create cognitive applications that support natural voice-based interactions with pre-existing webpages containing structured data such as tables and plots. Moreover, \textit{Reagent} readily learns vocabulary by asking questions when it doesn't understand the user's or the webpage's terminology. One avenue for further research is to improve the knowledge acquisition process by identifying domains of third-party websites and/or users that are likely to share a common vocabulary. Another is to extend the pointing capabilities to serve multiple users sharing a common workspace. The cognitive applications that we are  building~\cite{kephart2018cognitive,farrell2016symbiotic} can run in a wide range of physical environments, from laptops to large cognitive rooms. While technologies such as HTC Vive and Remote Mouse facilitate interactions in large-scale environments, leveraging standard mouse event capture capabilities in \textit{Reagent} currently limits it to a single user. To support multiple people pointing simultaneously, we plan to explore alternative pointing and gesture recognition approaches.

\bibliography{peveler}
\bibliographystyle{aaai}

\end{document}